\documentclass[aps,prb,
twocolumn,groupedaddress,superscriptaddress,showpacs]{revtex4}

\usepackage{amsmath}
\usepackage{graphicx}
\usepackage{hyperref}
 
\DeclareMathOperator{\sign}{sign}

\newcommand{\Hcal}{\mathcal{H}}

\newcommand{\tp}{t_{\perp}}

\newcommand{\om}{\omega}
\newcommand{\Om}{\Omega}

\newcommand{\vk}{{\bf k}}

\newcommand{\vq}{{\bf q}}
\newcommand{\vp}{{\bf p}}

\newcommand{\la}{\langle}
\newcommand{\ra}{\rangle}

\newcommand{\vf}{\ensuremath{v_\text{F\,}}}

\newcommand{\vv}{\bar{v}}
\newcommand{\vs}{\bar{\psi}}
\newcommand{\BB}{\rm BB}

\newcommand{\VMF}{V_{\text{MF}}}

\newcommand{\rA}{\text{A}}
\newcommand{\rB}{\text{B}}

\newcommand{\GR}{{\rm G}}

\begin{document}


\title{Transmission through a biased graphene bilayer barrier}

\author{Johan Nilsson}
\affiliation{Department of Physics, Boston University, 590 
Commonwealth Avenue, Boston, MA 02215, USA}

\author{A.~H. Castro Neto}
\affiliation{Department of Physics, Boston University, 590 
Commonwealth Avenue, Boston, MA 02215, USA}

\author{F. Guinea}
\affiliation{Instituto de  Ciencia de Materiales de Madrid, CSIC,
 Cantoblanco E28049 Madrid, Spain}

\author{N.~M.~R. Peres}
\affiliation{Center of Physics and Departamento de F{\'\i}sica,
Universidade do Minho, P-4710-057, Braga, Portugal}

\date{September 17, 2007}



\begin{abstract}
We study the electronic transmission through a graphene bilayer in the
presence of an applied bias between layers. 
We consider different geometries involving interfaces between
both a monolayer and a bilayer and between two bilayers.
The applied bias opens a sizable gap in the spectrum inside the
bilayer barrier region,  
thus leading to large changes in the transmission probability and 
electronic conductance that are controlled by the applied bias.

\end{abstract}

\maketitle


\section{Introduction.}
The idea of carbon based electronics has been around since the
discovery of carbon nanotubes almost fifteen years ago. Much progress
has been made but many problems associated with manufacturability
remain still to be resolved --
see for example the recent reviews in
Refs.~\onlinecite{nanotubereview2,nanotubereview3}.
Recently, another possible platform for carbon based electronics 
was discovered in graphene,\cite{Novoselov2004}  i.e., a
two-dimensional (2D) honeycomb 
lattice of carbon atoms that can be viewed either as 
a single layer of graphite or an unrolled nanotube.
The electric field effect has already been demonstrated in these systems:
by tuning a gate bias voltage one can control both the type (electrons or
holes) and the number of carriers.\cite{Novoselov2005,Zhang2005}
For a recent review on the rise of graphene see 
Ref.~\onlinecite{naturematrev}.

One fundamental difficulty with most of 
the graphene devices studied so far
is the experimental fact that there exists a universal (sample
independent)  maximum in the resistivity  of the order of 
$6.5 \text{k}\Om$ near the Dirac point in all these 
systems.\cite{Novoselov2005}
This relatively small resistivity limits the
performance of devices via a poor on-off ratio.
The reason for the nonzero minimal conductivity is the
peculiar gapless spectrum and presence of disorder in the samples.
Several theoretical studies using different methods
find a universal minimum in the conductivity in some limits. 
But the reason for it and its exact value 
varies,\cite{nuno2006_long,Katsnelson06a,Beenakker06,
  Nilsson2006b,Katsnelson06c,Nomura06,Snyman2007,Cserti2007}  
see also the recent review Ref.~\onlinecite{Antonio_Rev2007}.
There has also been a number of earlier theoretical studies of junctions
and barriers in both monolayer and bilayer graphene 
systems.\cite{Cheianov06,Katsnelson06b,Beenakker06,Pereira2006,
  Efetov06,Snyman2007} 

One way of getting around the minimal conductivity is to use nano-ribbons
made out of graphene. Because of the confinement these systems are 
generally found to be gapped,\cite{Son2007} 
which is consistent with recent experiments.\cite{Han2007}  
In this paper we propose another simple geometry that transforms the
semi-metallic graphene into a semiconductor with a truly gapped
spectrum without confinement.
The presence of a gap makes the properties of the system more
robust to perturbations. This is crucial for device performance
and possible device integration since imperfections are always present.
The basic idea is to use a bilayer region as a barrier for the
electrons. By manipulating the electrostatics with gates
and/or chemical doping a gap can appear in the spectrum.
By tuning the chemical potential one can move the system from
sitting inside of the gap, where the electronic transmission
is exponentially suppressed, into
the allowed band regions where the transmission is close to one.
Other interesting features of the proposed geometry are that 
the gap size can be tuned with gates, 
allowing for an external control of the electronic 
properties, and the bilayer can be integrated as one of the 
components of a pure graphene based device.\cite{berger04}
Recently barriers  made out of double-gated bilayer
graphene have been fabricated and characterized in 
Ref.~\onlinecite{Oostinga2007}, and their results are consistent
with a gate-tunable gap.

The paper is organized as follows:
In Section~\ref{sec:model} we introduce the model of the
system that we are considering and give explicit expressions
for the wave functions.
In Section~\ref{sec:geometries} we match wave functions
considering different geometries.
In particular we study a monolayer-bilayer interface, a
bilayer-bilayer interface, and two
bilayer barrier setups. In the barrier setups we use either
monolayer graphene or bilayer graphene in the leads.
In Section~\ref{sec:conductance} we present results for the
conductance in the different barrier setups. 
A brief summary and the conclusions of the paper
 can be found in  Section~\ref{sec:summary}.
For completeness
we also provide some details about the self-consistent determination
of the gap in Appendix~\ref{app:gap_self_consistent}.
In Appendix~\ref{app:matrices} we provide some details of how
we compute the transmission amplitudes.

\section{Model}
\label{sec:model}
\begin{figure}[htb]
\includegraphics[scale=0.28]{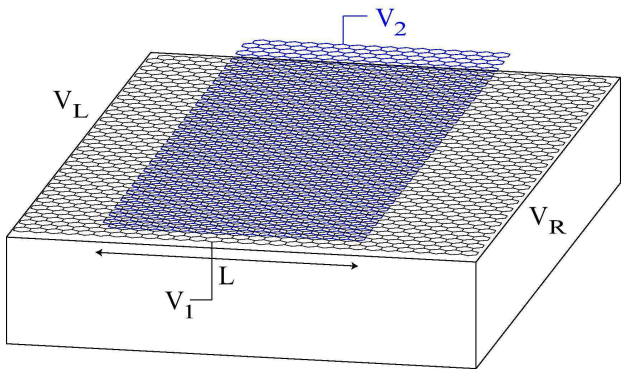}
\caption{[color online] Geometry of the biased graphene bilayer
  barrier with monolayer leads. }
\label{fig_device}
\end{figure}

A schematic picture of the system is shown in Fig.~\ref{fig_device}.
The basic structure involves a graphene sheet extending to the left
and to the right of a region where there is a second graphene sheet sitting on
top of the first, this region is the {\it bilayer barrier} (BB) region.
The whole structure is assumed to sit on top of a dielectric
spacer insulating the graphene from a back gate.
Later on (Sec.~\ref{sec:bilayeronly}) we will also consider a system where
the regions to the left and right of the barrier are made
out of graphene bilayers. We will refer to the two systems as 
having monolayer and bilayer leads respectively.

To the left ($L$) and to the right ($R$) of the barrier
the low-energy effective Hamiltonian (near the K point of the Brillouin
zone) has the form of the 2D Dirac Hamiltonian:\cite{Wallace47}
\begin{equation}
\label{eq_BBB:Hleft}
\Hcal_{\alpha} = \vf
  \begin{pmatrix}
    V_{\alpha} / \vf & k_x + i k_y \\
    k_x - i k_y & V_{\alpha} / \vf
   \end{pmatrix},
\end{equation}
where $\vk = (k_x,k_y)$ is the 2D
momentum measured relative to the K point, $\alpha = L, \, R$
and $\vf \approx 10^6 \text{m/s}$ is the Fermi-Dirac velocity 
(we use units such that $\hbar = 1 = \vf$ from now on).
The Dirac Hamiltonian acts on a spinor representing the
wave functions on the two sublattices: 
$\psi = (\psi_{\rA 1}, \, \psi_{\rB 1})^{T}$. 
In the following, we mostly work with the
case where $V_L = V_R = 0$ so that the energy is measured with respect to
the Dirac point in the monolayer regions. The spectrum of the Hamiltonian
in Eq.~(\ref{eq_BBB:Hleft}) is then simply $E_{\pm}(k) = \pm   k$, 
($k^2 = k_x^2+k_y^2$), where the plus (minus) sign is associated
with electron (hole) states.
The low-energy effective bilayer Hamiltonian has the form 
[see e.g. Refs.~\onlinecite{Paco06a,McCann2006a}]: 
\begin{equation}
  \label{eq_BBB:HBB}
  \Hcal_{\BB} =  
  \begin{pmatrix}
    V_1 & q_x + i q_y & \tp & 0 \\
    q_x - i q_y & V_1 & 0 & 0 \\
    \tp & 0 & V_2  & q_x - i q_y \\
    0 & 0 & q_x + i q_y & V_2 
  \end{pmatrix}.
\end{equation}
Here the two-dimensional momentum is $\vq = (q_x,q_y)$ and the corresponding
spinor is
$\psi = (\psi_{\rA 1}, \, \psi_{\rB 1}, \, \psi_{\rA 2}, \,
\psi_{\rB 2})^{T}$. $\tp \approx 0.35 \, \text{eV}$ 
is the hopping energy between nearest neighbor
atoms in different planes (i.e., $\rA 1$ and $\rA 2$). 
The monolayer is connected to plane 1 in the bilayer.
Solving for the spectrum of the Hamiltonian in
Eq.~(\ref{eq_BBB:HBB}) one finds four energy bands given by
\begin{multline}
  \label{eq:1}
E_{\pm,s}(q) = (V_1+V_2)/2  \\
\pm \sqrt{ q^2 + \frac{V^2}{4} + \frac{\tp^2}{2} 
  +s \frac{1}{2}\sqrt{4 ( V^2 +\tp^2) q^2 + \tp^4 }},
\end{multline}
where $s = \pm$ and $V = V_1-V_2$.
Thus for the two bands closest to the Dirac point ($E_{\pm,-}$)
the spectrum is gapped and has an unusual ``Mexican hat''
dispersion as was pointed out in Ref.~\onlinecite{Paco06a}.
There exists other examples in the literature of materials with a
similar dispersion, often called ``camel-back''-dispersion instead
of ``Mexican hat''. This feature can arise -- like in our case -- 
in the $\vk \cdot \vp$ approximation when two bands that are close to
each other in energy are allowed to hybridize. 
Examples of materials where a camel-back has been proposed include
Tellurium,\cite{tellurium2} 
GaP,\cite{camelbackGaP1978} 
and GaAs.\cite{camelbackAlAs2004}


The {\it crucial property} for the structure proposed here is that 
we allow for different voltages
on the two layers in the BB region: $V_1 \neq V_2$.
This possibility have been noted before,\cite{Falko2006a,Paco06a}
but here we are exploiting this feature.
These potentials can be created by a uniform electric field
through the bilayer that generates a charge imbalance between the
two layers. 
In a transport measurement the graphene is connected to 
electron reservoirs and can hence become charged.
When voltages are applied to the gates and the graphene
structure charge is redistributed to 
minimize the total electrostatic energy.
The problem is basically that of a capacitor. 
Upon studying the problem the bilayer must be
viewed as a single unit since the planes are connected by orbital overlap.
For example, the induced charge imbalance between the layers will screen the
applied electric field; $\tp$ also works against the applied field
since it tends to equalize the densities in the two planes. 

A simple estimate of these effects is provided by a self-consistent
Hartree theory [see Appendix~\ref{app:gap_self_consistent}
and Refs.~\onlinecite{McCann2006a,MacDonald_bilayergap_2006}]. 
For an isolated uncharged infinite bilayer 
we find that the net effect is to replace the 
applied voltage difference $V$
by a smaller effective $V_{\text{MF}}$.
For some reasonable parameters 
[$V < \tp$ and interlayer distance 
$d \approx 3.35-3.6 \, \text{\AA}$~\cite{graphitereview1,lundqvist06}],
$V_{\text{MF}}$ is down by a factor of order three 
compared with the bare value. 
In particular for an 
experimentally accessible voltage drop of $90 \, \text{V}$ over 
$300 \, \text{nm}$ we find an effective voltage difference of 
$V_{\text{MF}} \sim 40 \, \text{meV}$.
This value might be improved upon incorporating a dielectric
(e.g., $\text{SiO}_2$) 
and/or using a thinner dielectric spacer.
Thus it is not unreasonable to have 
$V_{\text{MF}} \sim 100 \, \text{meV}$.
This estimate was done before we became aware of the measurements reported
in Refs.~\onlinecite{Ohta06} and~\onlinecite{Castro06}.
In the first of these references the gap is measured in ARPES to be as large
as $200 \, \text{meV}$, but the charge densities are also quite large
in their case ($n \sim 1-6 \cdot 10^{13}  \, \text{cm}^{-2}$).
In the second reference the maximum obtained gap is estimated to
be $\sim 100 \, \text{meV}$. 
The largest possible value of the gap is estimated to be
$\sim 300 \, \text{meV}$ and is limited by the 
dielectric breakdown of the $\text{SiO}_2$.

Experimentally the bias can be controlled by different methods.
The conceptually simplest and most flexible method is to use a back
gate and a top gate (like in a dual gate MOSFET geometry), 
preferably using split gates and therefore allowing for different gate
voltages in the monolayer and bilayer regions. 
It is worth to mention that local top gates have already
been successfully fabricated on single-layer graphene 
samples.\cite{Lemme2007,Goldhaber2007a,Kim2007a,marcus2007a}
The field is progressing rapidly, since for example only a year ago
no top gate had been reported in graphene systems.
Moreover, very recently a top gate was also fabricated on bilayer 
graphene and characteristics consistent with a gate-tunable gap were
reported.\cite{Oostinga2007}
Another possibility is to change the chemical environment by depositing
donor or acceptor molecules on top of the structure.\cite{Ohta06,Castro06}
These act like dopants in a semiconductor and allow for independent
control of the bias and the chemical potential. Note however that
this method always introduces impurities into the system with
potentially important consequences.\cite{Nilsson2007}

On symmetry grounds a more general Hamiltonian than the one in
Eq.~(\ref{eq_BBB:HBB}) is certainly  allowed as discussed in
Ref.~\onlinecite{SW58} for the case of graphite.
For example, the couplings $\gamma_3$ and $\gamma_4$
associated with electron hopping between carbon atoms that are not nearest 
neighbors in different layers, familiar to the graphite literature,
are possible.\cite{graphitereview1}
In the BB the effects of these terms are less important than for the
low-energy features in graphite since the gap is a robust feature
in the low-energy spectrum of the BB. 
The electrostatic response in different sublattices within each layer
is also likely to be different, that is, one should use 
$V_{\rA 1} \neq V_{\rB 1}$ etc. 
If the applied voltage difference is not too large compared with $\tp$
the implications of this effect are presumably small.
%
The applied field and the pressure from anything on the top of
the structure will probably affect both the interlayer distance 
and the interlayer coupling.
The details of the band structure including all the effects
above is a very complicated problem that has yet to be studied.
Nevertheless, we believe that the Hamiltonian in 
Eq.~(\ref{eq_BBB:HBB}) correctly
captures the main features of the spectrum, including the important
formation of the gap in the spectrum near the Dirac point. 
In view of the large uncertainty in the parameters involved, it is
meaningless to pursue a more complicated model at this point.
The inclusion of the other parameters introduce no principal problems,
but the analysis becomes more complicated.

\subsection{The eigenvectors}
\label{sec:transmission_vectors}

The normalized energy $E$  eigenvectors of $\Hcal_{\alpha}$ in
Eq.~(\ref{eq_BBB:Hleft}) can be written as: 
\begin{equation}
  \label{eq_BBB:graphenevector}
\vv_{\alpha} 
=
\frac{1}{\sqrt{2} (E-V_{\alpha})}
\begin{pmatrix}
  E-V_{\alpha} \\ k_x - i k_y
\end{pmatrix},
\end{equation}
when the particles are ``on-shell'':
\begin{equation}
  \label{eq_BBB:dispk}
  (E-V_{\alpha})^2 = k_x^2 + k_y^2.
\end{equation}
A simple way to generate the eigenvectors of $\Hcal_{\BB}$ in
Eq.~(\ref{eq_BBB:HBB}) is to note that the columns of the 
Green's function $\GR^0$ in the bilayer region
($\GR^0  = \bigl[  E - \Hcal_{\BB} \bigr]^{-1}$)
are proportional to the eigenvectors if the relation
between the energy $E$ and the momentum $\vq$ are
``on-shell''. It is straightforward albeit
tedious to generalize this approach to the more general Hamiltonians
discussed above.
Using this we extract two different
energy $E$ eigenvectors of $\Hcal_{\BB}$ as
\begin{equation}
\label{eq_BBB:BBvector1}
\vv_{\BB,\rA 1} = 
\begin{pmatrix}
  \bigl[(E-V_2)^2 - q_x^2 - q_y^2 \bigr] (E -V_1) \\
  \bigl[(E-V_2)^2 - q_x^2 - q_y^2 \bigr]  (q_x - i q_y ) \\
  \tp (E-V_2) (E-V_1) \\ 
  \tp (E-V_1) (q_x + i q_y)
\end{pmatrix},
\end{equation}
and
\begin{equation}
\label{eq_BBB:BBvector2}
\vv_{\BB,\rA 2} = 
\begin{pmatrix}
  \tp (E-V_2) (E-V_1) \\ 
  \tp (E-V_2) (q_x - i q_y) \\ 
  \bigl[(E-V_1)^2 - q_x^2 - q_y^2 \bigr] (E -V_2) \\
  \bigl[(E-V_1)^2 - q_x^2 - q_y^2 \bigr]  (q_x + i q_y )
\end{pmatrix}.
\end{equation}
These just differ by their overall normalization.
It is straightforward to check
that these vectors are indeed eigenvectors of  $\Hcal_{\BB}$ in
Eq.~(\ref{eq_BBB:HBB}) by direct substitution.
The ``on-shell'' condition in the bilayer region reads:
\begin{multline}
  \label{eq_BBB:dispq}
  2(q_x^2 + q_y^2)  = (E - V_1)^2 + (E - V_2)^2
  \\
  \pm \sqrt{ \bigl[ (E - V_1)^2 - (E - V_2 )^2 \bigr]^2 
    + 4 \tp^2 (E - V_1) (E - V_2)}.
\end{multline}

\section{Different geometries}
\label{sec:geometries}
In this section we compute the transmission amplitudes for different edges
and geometries. By convention the incident wave is taken to 
arrive from the left side of the barrier and is transmitted to the
 right.

\subsection{Zig-zag termination with monolayer leads}
With our conventions the zig-zag termination of the barrier
corresponds to cutting the strip along the
y-direction.
For simplicity we consider the case that the width $W$ of
the structure is large 
enough so that the boundary conditions in the transverse direction
are irrelevant. It is then convenient to assume periodic boundary
conditions and use translational invariance and fix $k_y =
q_y$ to be a good quantum number in addition to the energy $E$.
This assumption can be relaxed.\cite{Nuno06a,Beenakker06,Efetov06}
When the width becomes small enough that the
quantization of the transverse momentum becomes important the
system becomes similar to a semiconducting nanotube with a finite radius. 
As discussed by Brey and Fertig, it is a good approximation to use the
continuum description of a graphene ribbon if it is wide enough and
provided that the proper boundary conditions for the continuous model are
employed.\cite{Brey06a}
For the zig-zag edges one can work in the single valley approximation
and the correct boundary condition is to take the
wave-function to vanish in one of the sublattices 
($\rA 2$ on the left and $\rB 2$ on the right in our case) 
at the BB boundaries.
We choose the energy $E$ and $k_y = |E|\sin(\phi)$ so that there are
propagating states to the left of the junction, with $k_x = E \cos(\phi)$.
The four solutions of Eq.~(\ref{eq_BBB:dispq}) for $q_x$ 
inside the bilayer we denote by $\pm q_{x1}$ and $\pm q_{x2}$.

\subsection{Monolayer-bilayer interface}
\begin{figure}[htb]
\includegraphics[scale=0.44]{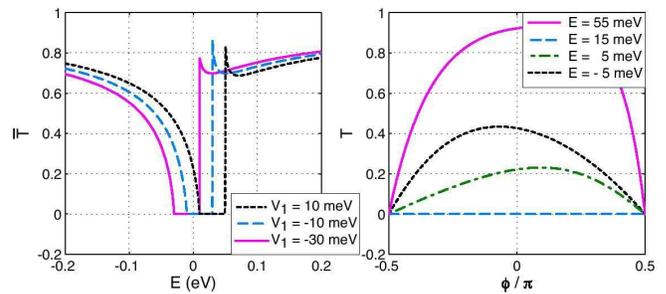}
\caption{[color online] Transmission amplitudes in the
  monolayer-bilayer step geometry.
  Left:  Energy dependence of $\overline{T}$ for 
  $V_2-V_1= 40 \, \text{meV}$ and different values of $V_1$.  
  Right: Angular dependence of $T(E,\phi)$ for 
  $V_1 = 10 \, \text{meV}$ and $V_2 = 50 \, \text{meV}$ and different values
  of the energy.}
\label{fig:zigzagstep}
\end{figure}

Consider the step geometry where there is no right end of
the bilayer. Then one must only keep states that propagate to the
right or decay as one moves into the bilayer.
Thus, to the left we take the wave function to be: 
\begin{equation}
  \label{eq_BBB:psileft1}
  \vs_L = 1 \vv_{L,+} e^{i k_x x} + r \vv_{L,-} e^{-i k_x x},
\end{equation}
where $r$ is the reflection amplitude, and $\vv_{L\pm}$ are the spinors
associated with the sublattices given in Eq.~(\ref{eq_BBB:graphenevector}). 
In the bilayer one has, 
\begin{equation}
  \label{eq_BBB:psitrans1}
  \vs_{\BB} = a_{1+} \vv_{\BB,1+} e^{i q_{x 1} x} 
  + a_{2+} \vv_{\BB,2+} e^{i q_{x 2} x},
\end{equation}
where $a_{1(2)\pm}$ are scattering amplitudes and $\vv_{\BB,1(2)\pm}$
the respective spinors computed from Eq.~(\ref{eq_BBB:BBvector1})
or Eq.~(\ref{eq_BBB:BBvector2}). 
Matching of the wave functions only involves their continuity.
Because the associated differential equation is of first order
this is sufficient to insure current conservation.
Explicitly the boundary conditions are:
\begin{subequations}
  \label{eq_BBB:Boundary_left}
\begin{eqnarray}
  \vs_L (x=0)\bigr|_{\rA 1} &=& \vs_{\BB}  (x=0)\bigr|_{\rA 1}, \\
  \vs_L (x=0)\bigr|_{\rB 1} &=& \vs_{\BB}  (x=0)\bigr|_{\rB 1}, \\
  \vs_{\BB} (x=0) \bigr|_{\rA 2} &=& 0.
\end{eqnarray}
\end{subequations}
From this we compute [for details see Appendix~\ref{app:matrices}]
the transmission probability 
$T(E,\phi) = 1 - |r|^2$ and the angular averaged transmission
probability
\begin{equation}
  \label{eq_BBB:Tbardef}
  \overline{T}(E) = \int_{-\pi/2}^{\pi/2} 
  \frac{d\phi}{ \pi} T(E,\phi).
\end{equation}
Some representative results are shown in Fig.~\ref{fig:zigzagstep}.
There are two features that are apparent in the figure:
(a) There is a small asymmetry between the angles $\pm \phi$. Hence, a
current of electrons without valley polarization leads to a transmitted
current with a finite valley polarization. This is not the case at other
boundaries, 
like a potential step applied to a graphene monolayer
or a graphene bilayer.\cite{Katsnelson06b}
The breaking of the symmetry between the two Dirac
points arises from the lack of time reversal symmetry, as we consider a
current carrying state, and the lack of inversion symmetry, induced by the
zigzag bilayer edge or the bias potential in the bilayer
(for a general discussion, see Ref.~\onlinecite{MGV07}, for a
particular discussion see Appendix~\ref{app:matrices}). As a result,
the barrier discussed here can be used as a device which creates a valley
polarized current.\cite{RTB07}
(b) There is a clear asymmetry between positive and negative
energies.
This can be understood by noting that the monolayer
is coupled to layer 1 in the bilayer.
When the energy is tuned to $V_1$ the weight on sublattice
$\rA 1$ goes to zero [c.f. Eq.~(\ref{eq_BBB:BBvector2})].
Consequently the current in plane 1 is zero at that energy,
and hence no current can flow into the bilayer. It appears
as though the current goes continuously to zero as the
band edge is approached.
At the other edge ($E \sim V_2$) of the band gap the current in 
plane 2 is zero  [c.f. Eq.~(\ref{eq_BBB:BBvector1})],
but the current can now flow in through the other plane.
This fact is responsible for the sharp edge in the transmission
amplitude at the conduction band edge.

It is important to choose the momenta for right-movers in
Eq.~(\ref{eq_BBB:psileft1}) and Eq.~(\ref{eq_BBB:psitrans1}) such that their
group velocity $v_{g} = dE/dq_{x} >0$. Otherwise one may erroneously
conclude that $T<0$ for some values of the energy, this is sometimes
referred to as the Klein paradox.\cite{Calogeracos99}
It is also worth noting that
the actual charge distribution near the edge is a complicated
problem that involves a self-consistent solution of the Poisson 
equation and the band structure, beyond the scope of this study.
This may lead to corrections to the simple wave function matching we
use here.
Furthermore, it is known that edges can lead to interesting effects in
graphene systems such as edge states and self-doping.\cite{nuno2006_long}

\subsection{Biased bilayer barrier with monolayer leads}
\begin{figure}[htb]
\includegraphics[scale=0.44]{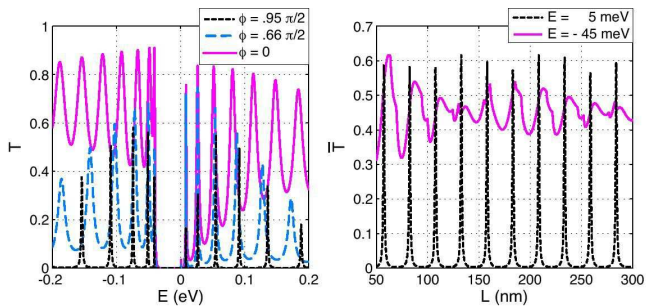}
\caption{[color online] Transmission amplitudes in the BB with
  monolayer leads.
Left: Energy dependence of $T(E,L,\phi)$ for different
  angles and $L=50 \, \text{nm}$.
  Right: Length dependence of $\overline{T}(E,L)$ for different
  energies. In both 
  the figures $V_1 = 0$ and $V_2 = -40 \, \text{meV}$.}
\label{fig:BBB_zigzag_1}
\end{figure}

Consider the barrier geometry of 
Fig.~\ref{fig_device}.
We assume the length $L$ of the bilayer region to be large compared to
the lattice spacing so that the continuum model is applicable. 
In this geometry one also needs the wave function to the right of the barrier:
\begin{equation}
  \vs_R = t \vv_{R,+} e^{i k_x (x-L)},
\end{equation}
where $t$ is now the transmission amplitude. 
Inside the barrier one generally needs all momentum components:
\begin{eqnarray}
\label{eq:bilayermiddle_vec}
  \vs_{\BB} &=& a_{1+} \vv_{\BB,1+} e^{i q_{x1} x} + a_{2+} \vv_{\BB,2+} e^{i
  q_{x2} x} \nonumber \\
&+&
a_{1-} \vv_{\BB,1-} e^{-i q_{x1} x} + a_{2-} \vv_{\BB,2-} e^{-i q_{x2} x}.
\end{eqnarray}
In this case, in addition to the boundary conditions in 
Eq.~(\ref{eq_BBB:Boundary_left})
there are also those at the right edge:
\begin{subequations}
  \label{eq_BBB:Boundary_right}
\begin{eqnarray}
  \vs_R (x=L)\bigr|_{\rA 1} &=& \vs_{\BB}  (x=L)\bigr|_{\rA 1}, \\
  \vs_R (x=L)\bigr|_{\rB 1} &=& \vs_{\BB}  (x=L)\bigr|_{\rB 1}, \\
  \vs_{\BB} (x=L) \bigr|_{\rB 2} &=& 0.
\end{eqnarray}
\end{subequations}
It is a simple task to match the boundary conditions and in this case
one finds six equations for the
six unknowns: $t$, $r$, $a_{1+}$, $a_{1-}$, $a_{2+}$, and $a_{2-}$.
The results for $T(E,L,\phi) = |t|^2$ 
for some representative parameters are shown in
Fig.~\ref{fig:BBB_zigzag_1}. 
%
%
The oscillations in the transmission amplitudes are due to the
possibility of having resonances inside of the BB region.
For example, in the right panel of Fig.~\ref{fig:BBB_zigzag_1} the 
two energies are chosen such that they have the same $q^2$ 
computed from Eq.~(\ref{eq_BBB:dispq}) for the propagating mode.
In one case ($E=5 \, \text{meV}$),
$|E|\sin(\phi) = q_y \ll q$ and the resonances are pronounced with the
distance between consecutive maxima approximately given by $\pi/q_x
\sim \pi/q \sim 25 \, \text{nm}$.
In the other case ($E=-45 \, \text{meV}$), $|E|\sin(\phi) = q_y \sim q$ 
and most of the resonance phenomena averages out 
upon performing the angle average because of the larger
variation of $q_x$.

\subsection{Armchair termination}
The calculations for the armchair geometry is similar albeit more
involved since one cannot work in the single cone 
approximation.\cite{Brey06a}
In this case, one must instead mix the valleys 
to be able to fulfill the boundary conditions so that the wave functions on
both of the lattice sites in layer 2 vanish at the boundaries. 
This doubles the size of the matrix problem that must be solved. 
For example, to compute the transmission
amplitudes in the BB geometry one must solve a system of twelve
equations. The calculation is a straightforward extension of the
case above and although the shapes of the curves are not exactly the
same there are no new features except for a parity
effect associated with the number of unit cells in the barrier.
This is related to the modulo 3 effect found in the spectrum of
an armchair graphene nano-ribbon in the nearest neighbor tight-binding
approximation.\cite{Brey06a}

\subsection{Bilayer-bilayer step and Biased bilayer barrier with bilayer leads}
\label{sec:bilayeronly}
It is important to note
that it is not necessary to have the BB region defined
by actually cutting the second graphene sheet.
Another possibility is to use a bilayer throughout, and to
use a local top gate to create a local gap and hence a
barrier in the bilayer.
As we shall see, the characteristics of this type of
junction is better than the one with the monolayer leads.
In particular the oscillations in the transmission
amplitudes and conductivities are much smaller because the
matching between two bilayers is usually better than
between a monolayer and a bilayer.

We will consider the simple case that the bilayer in the
$L$ and $R$ regions is essentially gapless,
we also assume that  
$|E| < \tp$ for the incoming wave as this will
likely be the experimental situation.
Since the gapless bilayer is a special case of the gapped
bilayer with no gap we may use the formulas and spinors from
Section~\ref{sec:transmission_vectors} directly with
$V_1 = V_2 = 0$. This case was considered 
 in Ref.~\onlinecite{Snyman2007} and allows for
some further simplifications of the spinors, but that is
not necessary here.

The incoming wave is taken to be 
a traveling wave with absolute momentum given by
$k^2 = |E| (\tp + |E|)$. As in the above we define
$k_x = \pm k \cos(\phi)$ and $k_y = k \sin(\phi)$.
One should also take care to define the sign of 
$k_x$ for the incoming wave so that the wave is a rightmover.
We will denote the corresponding spinors by $\vv_{\rB 0,\alpha}$, where
$\alpha = \pm$ goes with $\exp(\pm i k_x x)$ and denotes right-
and left-movers respectively.
To be able to fulfill the boundary conditions one must also
consider the decaying modes,\cite{Katsnelson06b} 
which will have an imaginary
value of the momentum in the $x$-direction: 
$k_x = i \kappa_x$. One can show that the correct value is
$\kappa_x = \sqrt{|E| (\tp - |E|) + k_y^2}$.
The corresponding spinors we write as $\vv_{\rB 0,i \alpha}$
in an obvious notation.

\begin{figure}[htb]
\includegraphics[scale=0.44]{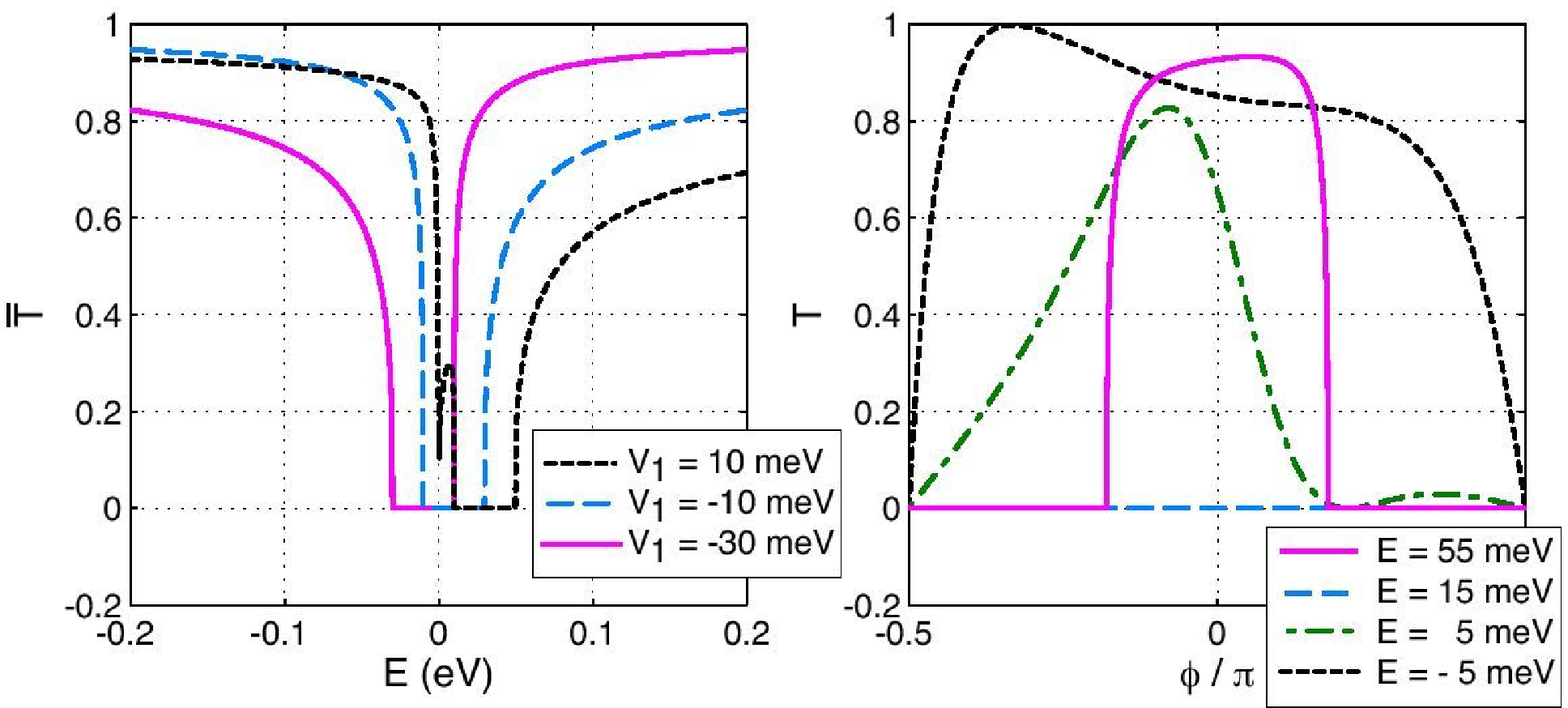}
\caption{[color online] Transmission amplitudes in the
  unbiased bilayer-biased bilayer step geometry.
  Left:  Energy dependence of $\overline{T}$ for 
  $V_2-V_1= 40 \, \text{meV}$ and different values of $V_1$.  
  Right: Angular dependence of $T(E,\phi)$ for 
  $V_1 = 10 \, \text{meV}$ and $V_2 = 50 \, \text{meV}$ and different values
  of the energy.}
\label{fig:bilayerstep}
\end{figure}

\begin{figure}[htb]
\includegraphics[scale=0.44]{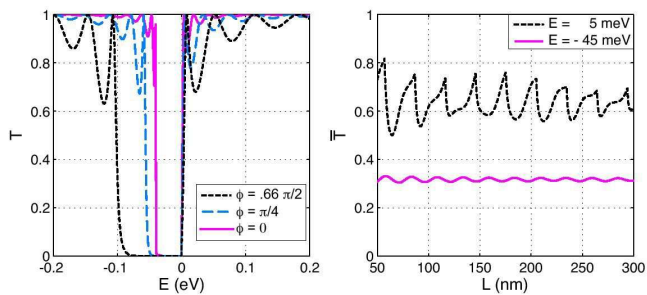}
\caption{[color online] Transmission amplitudes in the BB with
  bilayer leads.
  Left: Energy dependence of $T(E,L,\phi)$ for different
  angles and $L=50 \, \text{nm}$.
  Right: Length dependence of $\overline{T}(E,L)$ for different
  energies. In both 
  the figures $V_1 = 0$ and $V_2 = -40 \, \text{meV}$.}
\label{fig:bilayerbilayer_trans}
\end{figure}

The calculation proceeds exactly as in the other cases when
one has identified the particular incoming propagating mode in the
bilayer to the left:
\begin{equation}
  \label{eq:bilayeronly_vec1}
    \vs_{L} = 1 \vv_{\rB 0,+} e^{i k_x x} + r \vv_{\rB 0,-} e^{-i k_x x}
    + r' \vv_{\rB 0, -i} e^{ \kappa_x x}.
\end{equation}
Now we can consider a step geometry where the incoming wave from the
unbiased bilayer propagates into a biased bilayer. In this case the
spinor in Eq.~(\ref{eq:bilayeronly_vec1}) should be matched with
the one in Eq.~(\ref{eq_BBB:psitrans1}) at $x=0$. More details for
this case are provided in Appendix~\ref{app:matrices}. Some
representative results are shown in Fig.~\ref{fig:bilayerstep}, and
these should be contrasted with the case of a monolayer-bilayer step in
Fig.~\ref{fig:zigzagstep}. Note that the asymmetry between $\pm \phi$
is also present in this case.
For the case of a bilayer barrier with bilayer leads the spinor
to the right is
\begin{equation}
  \label{eq:bilayeronly_vec2}
    \vs_{R} = t \, \vv_{\rB 0,+} e^{i k_x (x-L)} 
    + t' \, \vv_{\rB 0, i} e^{ - \kappa_x (x-L)}.
\end{equation}
The wave functions in Eq.~(\ref{eq:bilayeronly_vec1})
and (\ref{eq:bilayeronly_vec2}) should be matched with the one
in Eq.~(\ref{eq:bilayermiddle_vec}) to the left and to the right.
In this case the correct boundary conditions is to take all of the
components of the 4-spinor wave function to be continuous at the two 
boundaries of the BB region.
The transmission amplitude is in this case given by
$T(E,\phi) = |t|^2$.
An example of the transmission amplitudes is shown
in Fig.~\ref{fig:bilayerbilayer_trans}. It is clear that the magnitude
of the oscillations in the amplitudes are much smaller than in the
cases involving monolayer leads.
Moreover, the pronounced resonances found in
the length-dependence of Fig.~\ref{fig:BBB_zigzag_1} 
is largely gone in this case.
This is probably due to the fact that the wave functions of two
bilayers are better matched than those of a monolayer and a bilayer.
It is also interesting to note that 
the effective gap becomes larger than the actual gap in the BB
for larger values of the angles.
This is due to the fact that, given the energy, the absolute value
of the momentum is much larger in the bilayer than the monolayer.
Consequently one has to go to larger values of the energy in
Eq.~(\ref{eq_BBB:dispq}) 
to have a mode that is not decaying inside of the BB region.

\section{Results for the conductance}
\label{sec:conductance}
\begin{figure}[htb]
\includegraphics[scale=0.43]{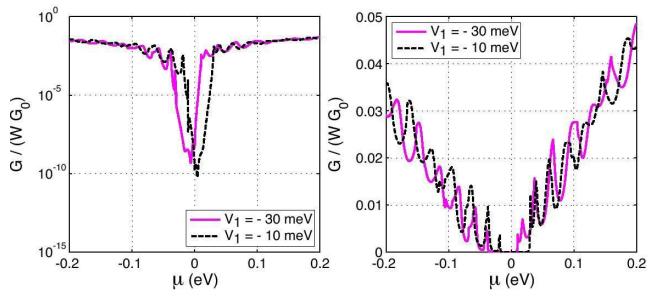}
\caption{[color online] 
  Left (right): Semi-log (linear) 
  plot of the zero-temperature
  conductance divided by the width $W$ (in nm) 
  of the BB device with monolayer leads
  in units of $G_0 = 4 e^2 /h$ as a function of the
  chemical potential $\mu$.
  $V_2-V_1= 40 \, \text{meV}$ and $L=50 \, \text{nm}$}
\label{fig:BBB_conductivity_1}
\end{figure}

\begin{figure}[htb]
\includegraphics[scale=0.43]{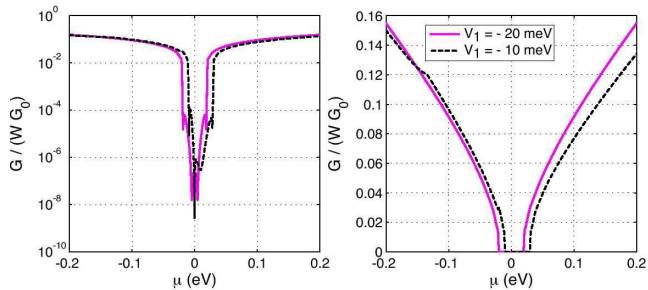}
\caption{[color online] Left (right): Semi-log (linear) 
  plot of the zero-temperature
  conductance divided by the width $W$ (in nm) 
  of the BB device with bilayer leads
  in units of $G_0 = 4 e^2 /h$ as a function of the
  chemical potential $\mu$.
  $V_2-V_1= 40 \, \text{meV}$ and $L=50 \, \text{nm}$}
\label{fig:BBB_conductivity_2}
\end{figure}

Using the Landauer formula [see e.g. Ref.~\onlinecite{DattaBook}], 
we find that the current across the BB is given by
\begin{equation}
  \label{eq_BBB:Landuaer1}
  I = \frac{2 e}{h} \int d E \sum_{n}
  |t_n(E)|^2 \bigl[  f(E-\mu_R) -f(E-\mu_L) \bigr].
\end{equation}
Here $f$ is the Fermi distribution function and 
$\mu_L$ ($\mu_R$) is the chemical potential in the
left (right) lead.
$n$ labels the modes and $t_n(E)$ the corresponding transmission
amplitude at energy $E$. 
From this the finite temperature linear response conductance can be computed
as a function of the overall chemical potential $\mu$:
\begin{equation}
  \label{eq_BBB:G_Tdep}
  G(\mu) = - \frac{4 e^2}{h} \int d E M(E) \overline{T}(E) 
  \frac{\partial f (E - \mu)}{\partial E}. 
\end{equation}
For the BB with monolayer leads,
$M(E) \sim W |E| / \pi$  is the number of transverse propagating
modes in the monolayer at energy $E$.
For bilayer leads the relation is instead 
$M(E) \sim W \sqrt{|E| (\tp + |E|)} / \pi$.
At zero temperature the expression in Eq.~(\ref{eq_BBB:G_Tdep})
simplifies to
\begin{equation}
  \label{eq_BBB:G_Tzero}
  G(\mu) = \frac{4 e^2}{h} M(\mu) \overline{T}(\mu). 
\end{equation}
Some results are presented in Fig.~\ref{fig:BBB_conductivity_1}
and Fig.~\ref{fig:BBB_conductivity_2}. 
As expected, if the
barrier is wide enough and the energy is tuned to be inside of the gap, 
the conductance is strongly suppressed.
Outside of this region the conductance is larger by many orders of
magnitude.
It is also clear from Eq.~(\ref{eq_BBB:G_Tdep}) that a finite temperature
$T$ will lead to a smearing of any sharp feature on an energy scale 
of approximately $4 \, T$. Nevertheless, as long as the temperature
is much smaller than the gap large on-off ratios are possible.

Let us finally comment on possible effects associated with roughness
or impurities at the edges of the sample. These will induce some
intervalley scattering and lead to an angle average.
This may be a serious problem for the proposals which emphasize
the angular dependence of the transmission.\cite{Katsnelson06b}
Because we are considering the transmission integrated over the
angles this should only weakly affect our results.
The resonances are also likely to survive in
a real sample when many incoming modes
overlap with one of the eigenmodes of the BB region.
Roughness at the ends of the BB will probably
broaden the resonances however. 

\section{Summary and conclusions}
\label{sec:summary}
We have studied the problem of electronic transmission through a graphene
bilayer barrier as a function of applied voltage between the layers
and the overall chemical potential.
We have considered two types of devices, one with monolayer leads and
one with bilayer leads. 
In the first type the barrier region is defined by having a bilayer
only in a small part of the sample. 
In the second type of device the barrier is instead defined by a local gate
in a system made entirely out of a graphene bilayer.
The latter system seems to have a smoother electronic characteristics
due to the absence of sharp boundaries that are present in the first device.
We have
shown that the transmission probability and the electronic conductance are
strongly dependent on the applied bias leading to the possibility of applying
this geometry for carbon based electronics.


\begin{acknowledgments}
We thank A. Geim for many illuminating discussions.
A.H.C.N. is supported through NSF grant DMR-0343790.
F.G. acknowledges funding from MEC (Spain) through grant
FIS2005-05478-C02-01, and the European Union contract 12881 (NEST).
N.M.R.P. acknowledges the financial support from POCI 2010
via project PTDC/FIS/64404/2006.
\end{acknowledgments}

\appendix

\section{Self-consistent determination of the gap}
\label{app:gap_self_consistent}
It is important to note that due to the polarization of the BB the
actual size of the potential difference $V$ between the planes is 
not equal to the bare externally applied potential difference $V_0$.
A simple approximation that takes into account the screening of the
external field by the BB is to include the interaction among the
electrons within the BB at the Hartree level in a self-consistent
manner.
Such a calculation was applied for the half-filled case for the
preprint of the present paper, and it was worked out independently 
at the same time for a more general case by McCann
in Ref.~\onlinecite{McCann2006a}.
More recent works along these lines includes a joint 
experimental-theory paper,\cite{Castro06} and
ab-initio calculations.\cite{MacDonald_bilayergap_2006}
There has also been a study of
the screening of an external electric field in graphene multilayer 
systems in the A-B stacking.\cite{Paco06b}

In this appendix we provide details of the self-consistent
determination of the gap. In
particular we give explicit analytic expressions for all of the
quantities involved in the non-linear equation that needs to be solved.
These expressions are potentially useful for anyone who wish to apply
this simple theory.

\subsection{Hartree Approximation}
\begin{figure}[htb]
\includegraphics[scale=0.43]{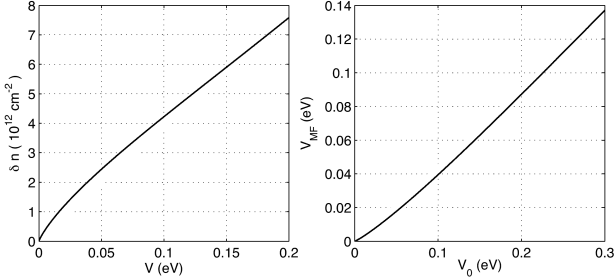}
\caption{Left:
Occupation asymmetry $\delta n$ at half filling as a function
of the bias $V$. 
Right: Self-consistently determined value of the bias potential
$\VMF$ as a function of the applied potential $V_0$.
}
\label{fig:selfconsistent}
\end{figure}

Charges on the two planes in the bilayer leads to an electrostatic
energy given by the capacitive coupling
\begin{equation}
  \label{eq:BGBbands_Ecapacitance}
  E_{c} = -2 \pi d e^2 S n_1 n_2,
\end{equation}
in the simplest approximation of two uniformly charged planes.
First we decouple the contributions from the total density 
$n = n_1 + n_2$ and the density difference $\delta n = n_2 - n_1$.
Here we are interested in the latter term,
which we will treat in the Hartree mean field approximation.
Denoting by $\la \delta n ( \VMF ) \ra$ 
the expectation values of $\delta n$ in the
ground state of the mean field Hamiltonian [i.e.,
Eq.~(\ref{eq_BBB:HBB}) with the substitution 
$V_1 - V_2 = V \rightarrow \VMF$], 
the mean field equation can be written as
\begin{equation}
  \label{eq:BGBbands_Mean_field1}
  \VMF = V_{0} - 2 \pi d e^2 \la \delta n ( \VMF ) \ra,
\end{equation}
where the expressions for $\la \delta n \ra$ at half-filling is 
given below in
Eq.~(\ref{eq:BGBbands_assymmetry_halffilling}).
If all the energies are expressed in eV the mean field equations
becomes simply $\VMF \approx V_0 - 7.3 \, \la \delta n ( \VMF ) \ra$.
The solution of the mean field equations at half-filling
is shown in Fig.~\ref{fig:selfconsistent}.
Away from half-filling one must also include the corrections due to
the partial filling of the conduction or valence band. 

An important quantity for the self-consistent determination of
the gap size is the density difference between the layers
$\delta n$. In the simple model we are using 
this quantity depends on $\tp$, $V$ and the density $n$.
At half-filling, when the chemical potential is sitting inside of the
gap the result to leading order in the cut-off (see below) is:
\begin{widetext}
\begin{multline}
  \label{eq:BGBbands_assymmetry_halffilling}
  \delta n =
\frac{V}{2 \pi (\tp^2 + V^2)^{3/2}}
\Biggl\{
V \sqrt{\tp^2 + V^2} \Bigl[ 
(V^2/2 + \tp^2 ) 
+ V \sqrt{\tp^2 + V^2/4})
\Bigr]
\\
 + \tp^2 (\tp^2 + V^2 / 2)
 \ln \Bigl[
 \frac{\tp^2 + V^2/2 + \sqrt{\tp^2 + V^2/4} \sqrt{\tp^2 + V^2}}
 {(\sqrt{\tp^2 + V^2} - V) V/2}
 \Bigr]
\Biggr\}.
\end{multline}
\end{widetext}
The occupation asymmetry $\delta n$ as a function of $V$ 
at half filling is also depicted in Fig.~\ref{fig:selfconsistent}.
One can easily check that the expression in 
Eq.~(\ref{eq:BGBbands_assymmetry_halffilling})
reduces to the correct expression in the limit of decoupled
planes ($\tp =0$).

\subsection{Occupation asymmetry at half-filling}
First we introduce the shorthand 
$ N_{\alpha j}/D \equiv \GR_{\alpha j \, \alpha j}^0$ 
for the diagonal components of the bare Green's function
that is defined by $\GR^ 0 = [\om - \Hcal_{\BB}]^{-1}$. 
We take $D = \text{Det}[\om - \Hcal_{\BB}]$, and
$N_{\alpha j}$ to be the appropriate cofactor of
the matrix $[\om - \Hcal_{\BB}]$.
Using this the density of states on sublattice $\alpha j$
can be written as
\begin{equation}
  \label{eq:BGBbands_rho1}
  \rho_{\alpha j}(\om) = \frac{1}{S} 
  \sum_{\beta = \pm}
  \sum_{\beta'=\pm}
  \sum_{\vk,\sigma} 
  \frac{N_{\alpha j}}{D'(E_{\beta,\beta'})} 
    \delta(\om - E_{\beta,\beta'}).
\end{equation}
Here the $\beta$-sums are over the different bands
and we have suppressed the frequency- and momentum-dependence
of the functions for brevity.
If we want to compute the density difference between the layers
$\delta n = n_2 - n_1$ when the chemical potential is sitting inside
the gap we must calculate
\begin{multline}
  \label{eq:BGBbands_assymmetry1}
  \delta n = \int_{-\infty}^{0} d\om
  \bigl[ \rho_{\rA 2} +\rho_{\rB 2}- \rho_{\rA 1} - \rho_{\rB 1} \bigr]
\\
= 4 \sum_{\beta=\pm} \int_0^{\Lambda} \frac{d^2\vk}{(2\pi)^2} 
\frac{ N_{\rA 2} + N_{\rB 2}- N_{\rA 1} -N_{\rB 1} }{D'(\om)}
\biggl|_{\om = E_{-,\beta}},
\end{multline}
which include both spin polarizations
 and the two valleys. Now one may use that
\begin{equation}
  \label{eq:BGBbands_numerator1}
  N_{\rA 2} + N_{\rB 2}- N_{\rA 1} -N_{\rB 1} 
  = -2 V ( \om^2 + k^2 - V^2/4 - \tp^2 /2 ),
\end{equation}
and
\begin{equation}
  \label{eq:BGBbands_denominator1}
  D'(E_{\pm,\beta}) 
  =  \beta 2 E_{\pm,\beta}\sqrt{4 (V^2 + \tp^2) k^2 + \tp^4},
\end{equation}
to write
\begin{multline}
  \label{eq:BGBbands_assymmetry2}
  \delta n = \frac{V}{\pi} \int_0^{\Lambda^2}
  \frac{d(k^2)}{\sqrt{4 (V^2 + \tp^2) k^2 + \tp^4}}
  \frac{E_{+,+} - E_{+,-}}{E_{+,+} \, E_{+,-}}
\\ \times
  \bigl[E_{+,+} \,  E_{+,-} - k^2 + V^2/4 + \tp^2 /2 \bigr].
\end{multline}
Using $E_{+,+} \, E_{+,-} = \sqrt{(k^2 - V/2)^2 + V^2 \tp^2/4}$
one can convince oneself that the integral in
Eq.~(\ref{eq:BGBbands_assymmetry2}) is convergent as 
$\Lambda \rightarrow \infty$ so that the leading term is independent
of the cutoff. Changing the integration variable to $z$ defined by
\begin{equation}
  \label{eq:BGBbands_zdef}
  z = \sqrt{4 (V^2 + \tp^2) k^2 + \tp^4},
\end{equation}
the integral can be performed analytically with the result
shown in Eq.~(\ref{eq:BGBbands_assymmetry_halffilling}).

\section{Explicit matrix equations}
\label{app:matrices}

In this appendix we show how to obtain the matrix equations
that we then solve numerically to obtain the transmission amplitudes.
For the simplest case of a monolayer--bilayer step,
using
Eqs.~(\ref{eq_BBB:psileft1}),~(\ref{eq_BBB:psitrans1}),~(\ref{eq_BBB:graphenevector}),
and~(\ref{eq_BBB:BBvector1}) 
the boundary conditions in Eq.~(\ref{eq_BBB:Boundary_left}) can
be rewritten as a matrix equation:
\begin{widetext}
\begin{equation}
  \label{eq:2}
  \begin{pmatrix}
    E \\ k_x - i k_y \\ 0
  \end{pmatrix}
  =
  \begin{pmatrix}
    E            & \bigl[(E-V_2)^2 - q_{x1}^2 - k_y^2 \bigr] (E -V_1)
                 & \bigl[(E-V_2)^2 - q_{x2}^2 - k_y^2 \bigr] (E -V_1) \\ 
    -k_x - i k_y  & \bigl[(E-V_2)^2 - q_{x1}^2 - k_y^2 \bigr]  (q_{x1} - i k_y )
                 & \bigl[(E-V_2)^2 - q_{x2}^2 - k_y^2 \bigr]  (q_{x2} - i k_y ) \\ 
    0            & \tp (E-V_2) (E-V_1)
                 & \tp (E-V_2) (E-V_1)
  \end{pmatrix}
  \begin{pmatrix}
    -r \\ a_{1+} \\ a_{2+}
  \end{pmatrix}.
\end{equation}
For this simple case it is also possible to work out
an explicit expression for $r$:
\begin{equation}
  \label{eq:3}
  r = - \frac{E^3-2 V_2
   E^2-\left[ k_y^2 + q_{x1}^2+ q_{x2}^2- V_{2}^2 + q_{x1}
   q_{x2} -k_{x} (q_{x1} + q_{x2} ) \right] E -( k_{x}-i
   k_{y} ) ( q_{x1} + q_{x2}) V_{1}}
   {E^3- 2 V_{2}
   E^2- \left[ k_{y}^2 + q_{x1}^2 + q_{x2}^2 - V_{2}^2 + q_{x1}
   q_{x2} + k_{x} ( q_{x1} + q_{x2} ) \right] E + ( k_{x}+i
   k_{y} ) ( q_{x1} + q_{x2} ) V_{1}}.
\end{equation}
\end{widetext}
Now one can substitute the correct values of the momenta
[c.f. Eq.~(\ref{eq_BBB:dispq})]
such that the incoming state is a right-mover and that
only states that decay or propagate to the right inside
of the bilayer are present.
We note that for energies such that $q_{x1}$ is real and
$q_{x2}$ is imaginary -- which is often the case for energies
such that $V \lesssim |E - (V_1 +V_2)/2| \lesssim \tp$ -- the transformation 
$k_y \rightarrow -k_y$ is in general not simply associated with a
phase of $r$. Consequently there is an asymmetry in the transmission
amplitude as shown in Fig.~\ref{fig:zigzagstep}.

The reason for the asymmetry is the broken inversion symmetry.
Either the symmetry is broken by a zig-zag edge of the bilayer
or by the bias field.
It is clear that the bias potential breaks the inversion
symmetry in the point in the middle between the two A atoms
of the unit cell. This is crucial as it breaks the sublattice
symmetry that is otherwise present.
It is also important that there exists a mode that is evanescent
as this breaks the symmetry between states with 
$k_y$ and $-k_y$ when one is matching the wave functions. 
For if all momenta are real,
one can easily convince oneself that to solve for
$-k_y$ one must only take the complex conjugate of
the solution with $k_y$, thus the only difference is the phase between
$r$ and $r^*$ which will not affect the transmission amplitude.
In addition it is necessary that there exists a mode that can
transmit the current. Otherwise all of the incoming current is
reflected and no asymmetry can be generated, note that this is
the case for a monolayer with a gap.

The other more complicated cases involves $6 \times 6$ and
$8 \times 8$ matrices. The procedure to obtain the transmission
amplitude is a straightforward generalization of the example
worked out above, but the full form of the matrices are too
long to write out here.
The matrix inversion is then performed numerically.

Another interesting example is to match a wave coming in from an 
unbiased bilayer and propagating into a biased bilayer.
In this case the matrix equation can be written as:
\begin{widetext}
\begin{multline}
  \label{eq:2}
  \begin{pmatrix}
    -E \\ -(k_x - i k_y) \\ |E| \\ \sign(E)(k_x + i k_y)
  \end{pmatrix}
  =
\\
  \begin{pmatrix}
    -E           &  E           
                 & \bigl[(E-V_2)^2 - q_{x1}^2 - k_y^2 \bigr] (E -V_1)
                 & \bigl[(E-V_2)^2 - q_{x2}^2 - k_y^2 \bigr] (E -V_1) \\ 
    -(-k_x - i k_y) & (-i \kappa_x - i k_y)  
                 & \bigl[(E-V_2)^2 - q_{x1}^2 - k_y^2 \bigr]  (q_{x1} - i k_y )
                 & \bigl[(E-V_2)^2 - q_{x2}^2 - k_y^2 \bigr]  (q_{x2} - i k_y ) \\ 
    |E|          & |E|
                 & \tp (E-V_1) (E-V_2)
                 & \tp (E-V_1) (E-V_2) \\
    \sign(E)(-k_x + i k_y) &  \sign(E)(-i \kappa_x + i k_y) 
                 &   \tp (E-V_1) (q_{x1} + i k_y) &   \tp (E-V_1) (q_{x2} + i k_y) 
  \end{pmatrix}
\\ \times
  \begin{pmatrix}
    -r \\ -r' \\ a_{1+} \\ a_{2+}
  \end{pmatrix}.
\end{multline}  
\end{widetext}
Here we have used the fact that the spinors simplify in the
leads where the inversion symmetry is not broken 
(i.e. $V_1 = V_2 = 0$).\cite{Snyman2007}
Also in this case one finds that there is an asymmetry between
negative and positive angles within each valley.
In this case this is due to the fact that the inversion symmetry is
broken by the bias potential in the BGB. In the case of 
$V_1 = V_2 \neq 0$ the inversion symmetry is not broken and
the transmission is symmetric between $\pm k_y$.\cite{Katsnelson06b}


\bibliography{graphite}

\end{document}